# ℓ -state solutions of a new four-parameter *1/r^2* singular radial non-conventional potential via asymptotic iteration method

A. J. Sous

Faculty of Technology and Applied Sciences,
Al-Quds Open University, Tulkarm, Palestine.

Email: asous@qou.edu

Abstract

In the present work, we give a numerical solution of the radial Schrödinger equation for new four-parameter radial non-conventional potential, which was introduced by Alhaidari. In our calculations, we applied the asymptotic iteration method (AIM) to calculate the eigenvalues of the potential for arbitrary parameters and any $l$ state. It is found that this method gives highly accurate results that compares favorably with other. Moreover, some new results were presented in this paper.

## 1. Introduction

There has been a growing interest in investigating the numerical solutions of the Schrödinger equation for some physical potential models. This is because exact solutions of the Schrödinger equation are very limited, and only obtainable for a few number of physical potentials [1-3]. The hyperbolic potentials are commonly used to model inter-atomic and intermolecular phenomena. Among such potentials are the Pöschl-Teller, Rosen-Morse , Scarf, and hyperbolic single wave potential, which have been studied extensively in the literatures [4-11]. The hyperbolic potential under investigation here is, in fact, a generalization of the hyperbolic Pöschl-Teller potential. This four-parameter radial potential was introduced by A. D. Alhaidari [2,3] and reads as follows:

$$V(r) = \frac{1}{\sinh^2(\lambda r)}[V_0 + V_1 \tanh^2(\lambda r) + V_2 \tanh^4(\lambda r)], \qquad (1)$$

Where $\{\lambda, V_i\}$ are real parameters such that $\lambda > 0$, $V_0 > 0$, $V_2 \neq 0$, and $\lambda$ is a length scale that determines the range of the potential. Near the origin, it is $1/r^2$ singular, but as $r \to \infty$ it decays exponentially to zero signifying that it is short-range.

There are three distinct physical configurations of the potential (1). The first one is when the potential has two local extrema (one local minimum and one local maximum). In this configuration, the potential could have resonances but no bound states, or it could have both. The second configuration occurs when the two extrema coincide at an inflection point or when the potential has no local extrema. In these two cases, the potential can support neither bound states nor resonances. In the third configuration, the potential has one local minimum and could support only bound states but no resonances. Due to the shortness of the potential range, we expect that the size of the bound states energy spectrum to be finite. These speculations will be verified below [2,3].

If $V_2 = 0$, potential (1) becomes the well-known hyperbolic Pöschl-Teller potential, which belongs to the conventional class of exactly solvable problems [2,3] and has the following energy spectrum formula:

$$E_n = -\frac{\lambda^2}{2}\left(2n+1+\sqrt{\tfrac{1}{4}+2V_0/\lambda^2}-\sqrt{\tfrac{1}{4}-2V_1/\lambda^2}\right)^2 \qquad (2)$$

where $n = 0,1,...,N$ and N is the largest integer less than or equal to $\tfrac{1}{2}\sqrt{\tfrac{1}{4}-2V_1/\lambda^2}-\tfrac{1}{2}\sqrt{\tfrac{1}{4}+2V_0/\lambda^2}-\tfrac{1}{2}$ [2,3].

For non-zero angular momentum, A. D. Alhaidari wrote the four-parameter radial potential as $V(r) = \frac{V_0}{r^2} + \tilde{V}(r)$, which makes $\tilde{V}(r)$ a non-singular short-range potential and gives the total Hamiltonian as

$$H = -\frac{\hbar^2}{2m}\frac{d}{dr^2} + \frac{\ell(\ell+1)+2V_0}{2r^2} + \tilde{V}(r) \qquad (3)$$

Thus, the time-independent radial Schrödinger's equation can be written as

$$\frac{-\hbar^2}{2m}\psi''(r) + \left[\frac{\ell(\ell+1)}{2r^2}+V(r)\right]\psi(r) = E\,\psi(r) \qquad (4)$$

In this paper, we present the Asymptotic Iteration Method (AIM) to calculate the eigenvalues not only for one-dimensional Schrödinger equation, but also for the three-dimensional and spherically symmetric radial case.

## 2. Basic Elements of the Asymptotic Iteration Method

The AIM method is based on solving a second order differential equation of the form:

$$f_n''(x) = \lambda_0(x)\,f_n'(x) + s_0(x)\,f_n(x) \qquad (5)$$

Where $\lambda_0 \neq 0$ and the prime denotes the derivative with respect to $x$. The variables, $s_0(x)$ and $\lambda_0(x)$ are sufficiently differentiable. To find a general solution to this equation, we differentiate (5) with respect to $x$ and find

$$f_n'''(x) = \lambda_1(x) f_n'(x) + s_1(x) f_n(x) \tag{6}$$

Where
$$\begin{aligned}\lambda_1(x) &= \lambda_0'(x) + s_0(x) + \lambda_0^2(x),\\ s_1(x) &= s_0'(x) + s_0(x) \lambda_0(x)\end{aligned} \tag{7}$$

Similarly, the second derivative of (5) yields

$$f_n^{(4)}(x) = \lambda_2(x) f_n'(x) + s_2(x) f_n(x) \tag{8}$$

Where

$$\begin{aligned}\lambda_2(x) &= \lambda_1'(x) + s_1(x) + \lambda_0(x) \lambda_1(x),\\ s_2(x) &= s_1'(x) + s_0(x) \lambda_1(x)\end{aligned} \tag{9}$$

Equation (5) can be easily iterated up $(k+1)th$ and $(k+2)th$ derivatives, $k = 1, 2, 3, \ldots$

Therefore, we have the recurrence relations

$$\begin{aligned}f_n^{(k+1)}(x) &= \lambda_{k-1}(x) f_n'(x) + s_{k-1}(x) f_n(x)\\ f_n^{(k+2)}(x) &= \lambda_k(x) f_n'(x) + s_k(x) f_n(x)\end{aligned} \tag{10}$$

Where

$$\begin{aligned}\lambda_k(x) &= \lambda_{k-1}'(x) + s_{k-1}(x) + \lambda_0(x) \lambda_{k-1}(x),\\ s_k(x) &= s_{k-1}'(x) + s_0(x) \lambda_{k-1}(x)\end{aligned} \tag{11}$$

From the ratio of the $(k+1)th$ and $(k+2)th$ derivatives, we have

$$\frac{d}{dx} \ln[f_n^{(k+1)}(x)] = \frac{f_n^{(k+2)}(x)}{f_n^{(k+1)}(x)} = \frac{\lambda_k(x)[f_n'(x) + \frac{s_k(x)}{\lambda_k(x)} f_n(x)]}{\lambda_{k-1}(x)[f_n'(x) + \frac{s_{k-1}(x)}{\lambda_{k-1}(x)} f_n(x)]} \tag{12}$$

For sufficiently large $k$, if

$$\frac{s_k(x)}{\lambda_k(x)} = \frac{s_{k-1}(x)}{\lambda_{k-1}(x)} = \alpha(x) \tag{13}$$

which is the "asymptotic" aspect of the method, then, (12) reduces to

$$\frac{d}{dx}\ln[f_n^{(k+1)}(x)] = \frac{f_n^{(k+2)}(x)}{f_n^{(k+1)}(x)} = \frac{\lambda_k(x)}{\lambda_{k-1}(x)} \qquad (14)$$

which yields the general solution of (5) can be obtained as:

$$f_n(x) = \exp(-\int^x \alpha(x)dx_1)[C_2 + C_1\int^x \exp(\int^{x_1}[\lambda_0(x_2) + 2\alpha(x_2)]dx_2)dx_1] \qquad (15)$$

The termination condition of the method in (13) can be arranged as

$$\Delta_k(x) = \lambda_k(x)\,s_{k-1}(x) - \lambda_{k-1}(x)\,s_k(x) \qquad (16)$$

where $k$ shows the iteration number. For the exactly solvable potentials, the energy eigenvalues are obtained from the roots of (16), and the eigenvalues will not depends to chose of $x$, and the radial quantum number $n$ is equal to the iteration number $k$ for this case. However, for nontrivial potentials that have no exact solutions and for a specific principal quantum number $n$, we choose a suitable $x_0$ point, determined generally as the maximum value of the asymptotic wave function or the minimum value of the potential [12-15]. Then, the approximate energy eigenvalues are obtained from the roots of Eq. (16) for sufficiently large number of iterations $k$, where $k$ is always greater than $n$.

The stability of the results in AIM depends on different factors: First; an appropriate choice of coordinate transformations and functional transformations to transfer Schrödinger equation into suitable form, because of the fact that choosing unsuitable coordinate transformations lead to instability results. Second; the stability of results depends on choosing a suitable $x_0$ point which leads to correct and stability results, and if we change $x_0$ point to another value, we might reach to instability results, that is why it is critical to choose the right initial value $x_0$ very carefully. Third; when choosing suitable $x_0$ point and substituting it in equation (16), then as much as we increase the iterations we can reach to convergence and stability results.

## 3. The Potential Eigenvalues by AIM.

In order to overcome the convergence problem noted above, we make a change of variables as follows $x = 2\tanh(\lambda r)^2 - 1$

After making the convenient change of variable, a straightforward calculation shows that equation (4) becomes

$$\frac{-h^2}{2m}\left(2\lambda^2 \phi^2(x)\,\omega(x)\,\frac{d}{dx^2} + (\lambda^2\phi(x)\,\chi(x))\,\frac{d}{d}\right)\psi + $$
$$(\lambda^2\frac{l(l+1)}{2\eta(x)} + V(x) - E)\psi = 0 \qquad (17)$$

Where $\omega(x) = (x+1)$, $\qquad (18)$

$$\phi(x) = (x-1) \qquad (19)$$

$$\chi(x) = 3x + 1 \qquad (20)$$

$$\eta(x) = \operatorname{arctanh}^2(\frac{1}{2}\sqrt{2x+2}) \qquad (21)$$

with $\quad V(x) = -\frac{\phi(x)}{\omega(x)}[V_0 + \frac{1}{2}V_1\,\omega(x) + \frac{1}{4}V_2\,\omega^2(x)] \qquad (22)$

then, we obtain the second-order homogeneous linear differential equation in the following form:

$$\frac{d^2\psi(x)}{dx^2} = \lambda_0(x)\frac{d\,\psi(x)}{dx} + s_0(x)\,\psi(x) \qquad (23)$$

Where $\lambda_0(x) = \frac{-1}{2}\frac{\chi(x)}{\phi(x)\,\omega(x)} \qquad (24)$

and

$$s_0(x) = \frac{-2m}{8h^2\lambda^2\phi(x)^2\,\omega^2(x)\eta(x)}[4\phi(x)\eta(x)V_0 + 2\phi(x)\omega(x)\eta(x)V_1 + \phi(x)\omega^2(x)\eta(x)V_2$$
$$+ 4\omega(x)\eta(x)E - 2\lambda^2\,l(l+1)\,\omega(x)\,] \qquad (25)$$

It is now possible to calculate $\lambda_k(x)$, and $s_k(x)$ applying equation (11). Finally, one finds the energy eigenvalues of the potential in (1) by using the quantization condition given in (16) and choosing the proper initial point $x_0$. In order to improve the energy eigenvalues, the iteration number has to be increased within the range of convergence of the method.

## 4. Results and Discussions

In Table 1, a comparison between AIM results and numerical results obtained by the tridiagonal representation approach (TRA) for the potential (1) with $V_0 = 1, V_1 = -50, V_2 = 2, \lambda = 1, l = 0$ ,and $\hbar = m = 1$. It is found that the results obtained by AIM are in good agreement with the results of the tridiagonal representation approach and up to 8 significant digits in the ground state.

In Table 2, we presented the eigenvalues for the potential (1) when $V_2 = 0$ which corresponds to the well-known and exactly solvable hyperbolic Pöschl-Teller potential. This is to illustrate the accuracy of the AIM as compared to the TRA. We took $V_0 = 1, V_1 = -50, V_2 = 0, \lambda = 1, l = 0$, and $\hbar = m = 1$. The Table shows that the eigenvalues calculated by AIM, are in a good agreement with the exact values obtained from equation (2).

In Table 3, we compare the bound states energies obtained by the AIM and those obtained by the complex scaling method in [2,3]. The parameters are taken as $V_0 = 2, V_1 = -80, V_2 = 120, \lambda = 1$ and for various values of the angular momentum $\ell$. Our results are in good agreement with those listed in [2,3].

In Table 4, we presented the eigenvalues for the potential (1), the results obtained with $V_0 = 0, V_1 = -70, V_2 = 20, \lambda = 1$, and various values of the angular momentum $l$. The results in this Table considered as new results which have not been addressed before, and it confirm that the AIM is valid for arbitrarily values of parameters.

In our calculation we have chose several values for $x_0$ and found the unique one that does not produce instability and that value was $x_0 = 0$. Finally, we point out that The accuracy of the results for the higher excited states could be increased if the number of iterations were increased limited only by the stability and convergence properties of the AIM.

## 5. Conclusion

We calculated the energy eigenvalues of the Schrödinger equation for a new four-parameter *1/r^2* singular non-conventional potential using the AIM. Our method is easy to apply and leads to a good agreement with the complex scaling method (TRA, also its agreement with the results obtained by tridiagonal representation approach (TRA). To the best of our knowledge, this paper is the first to study the eigenvalues associated with this new four-parameter radial non-conventional potential.

## Acknowledgements

I would like to thank Prof. A. D. Alhaidari for his comments, and for his fruitful discussions and valuable suggestions on this written work. Many thanks also to Dr. Eng. Islam Amro Director of (ICTC), and to Dr. Bassam Al Turk and all technical staff in (ICTC) for their assistant during the research.

## References

[1] H. Bahlouli and A. D. Alhaidari, *Extending the class of solvable potentials: III. The hyperbolic single wave*, Phys. Scr. **81** (2010) 025008

[2] A. D. Alhaidari, *Solution of the nonrelativistic wave equation using the tridiagonal representation approach*, J. Math. Phys. Vol. **58** (2017) 072104

**[3]** A. D. Alhaidari, *Extending the class of solvable potentials. IV Inverse square potential with a rich spectrum,* **https://arxiv.org/abs/1706.09212**

[4] A D Alhaidari, H. Bahlouli, *Two New Solvable Potentials*, Journal of Physics A: Mathematical and Theoretical, 42, No. 26 (2009)

[5] AbdallahJ. Sous, *The Asymptotic Iteration Method for the Eigenenergies of the a Novel Hyperbolic Single Wave Potential*, Journal of Applied Mathematics and Physics, Vol.03 No.11(2015)

[6] D. Agboola, *Solutions to the Modified Pöschl–Teller Potential in D-Dimensions*, Chinese Physics Letters 27(4), (2010) 040301

[7] Chun-Sheng Jia, Tao Chen, Li-Gong Cui, *Approximate analytical solutions of the Dirac equation with the generalized Pöschl–Teller potential including the pseudo-centrifugal term,* Physics Letters A, (2009). 1621

[8] Gao-Feng Wei ,Shi-Hai Dong , *The spin symmetry for deformed generalized Pöschl–Teller potential,* Physics Letters A, 373, (2009) 2428

[9] Miloslav Znojil , *pT -symmetrically regularized Eckart, Pöschl-Teller and Hulthén potentials*, Journal of Physics A: Mathematical and General, 33, (2009)

[10] ] A. D. Alhaidari, *Solution of the nonrelativistic wave equation using the tridiagonal representation approach*, Journal of Mathematical Physics , 58, ( 2017) 072104

[11] Hassanabadi Hassan, Yazarloo Bentol Hoda, Liang Liang Lu, *Approximate Analytical Solutions to the Generalized Pöschl—Teller Potential in D Dimensions*, Chin.Phys.Lett. 29 (2012) 020303

[12] Hakan Ciftci, Richard L. Hall, Nasser Saad, *Asymptotic iteration method for eigenvalue problems,* J. Phys.A 36, (2003) 11807

[13] T.Barakat, *The asymptotic iteration method for the eigenenergies of the anharmonic oscillator potential*, Physics Letters A, (2005) 411

[14] Abdallah. J. Sous , *The Asymptotic Iteration Method for theEigenenergies of the a Novel Hyperbolic Single Wave Potential*, Journal of Applied Mathematics and Physics, 3 , (2015) 1406

[15] Abdullah .J. Sous, Abdulaziz D. Alhaidari, *Energy Spectrum for a Short-Range 1/r Singular Potential with a Non-Orbital barrier Using the Asymptotic Iteration Method*, Journal of Applied Mathematics and Physics, 04, (2016)

**Table1.** A comparison of the energy eigenvalues $E_n$ of the potential (1) with $\hbar = m = 1$ and $V_0 = 1, V_1 = -50, V_2 = 2, \lambda = 1, l = 0$.

| n | $E_n$ by TR Approach | $E_n$ by AIM |
|---|---|---|
| 0 | -27.878950096074 | -27.87895010 |
| 1 | -14.799140053574 | -14.79914003 |
| 2 | -5.854541479288 | -5.854537386 |
| 3 | -0.996376819225 | -1.003141164 |

**Table 2.** A comparison of the exact energy eigenvalues $E_n$ of the potential (1) with $V_2 = 0$ as obtained by the exact formula of Eq. (2) and those obtained numerically by the AIM and the TRA. We took $\hbar = m = 1$, and $V_0 = 1, V_1 = -50, V_2 = 0, \lambda = 1, l = 0$.

| n | $E_n$ Exact (Eq. 3) | $E_n$ by AIM | $E_n$ by TRA |
|---|---|---|---|
| 0 | -28.21876953 | -28.21876951 | -28.21876951 |
| 1 | -15.19378513 | -15.19378521 | -15.19378511 |
| 2 | -6.168800730 | -6.168809024 | -6.16880072 |
| 3 | -1.143816328 | -1.152171723 | -1.14394908 |

**Table 3**. A comparison of the energy eigenvalues $E_n$ of the potential (1) obtained here by the AIM and compared to those obtained by the complex scaling Method (CSM) in [5]. We took $V_0 = 2, V_1 = -80, V_2 = 120, \lambda = 1$ and for various values of the angular momentum L.

| $\ell$ | n | $E_n$ by CSM in [5] | $E_n$ by AIM |
|---|---|---|---|
| 0 | 0 | - 27.66703017245 | -27.66215332 |
|   | 1 | -4.96995355885 | -4.962786443 |
| 1 | 0 | -21.21593606495 | -21.09575480 |
|   | 1 | -0.8517865495 | -.7002047775 |
| 2 | 0 | -11.585302647445 | -11.56852380 |
| 3 | 0 | -1.44701935596 | -1.448553820 |

**Table 4.** The energy eigenvalues $E_n$ of the potential (1) with $V_0 = 0, V_1 = -70, V_2 = 20, \lambda = 1$, and various values of the angular momentum L.

| n | L=0 | L=1 | L=2 | L=3 |
|---|---|---|---|---|
| 0 | -63.61657472 | -40.32439957 | -30.00145387 | -20.83425508 |
| 1 | -40.74048413 | -22.75675584 | -15.04669318 | -8.687891472 |
| 2 | -23.13743830 | -10.35480424 | -5.271906619 | -1.615752588 |
| 3 | -10.67884685 | -2.918758897 | -0.6798685034 | |
| 4 | -3.122016663 | -0.3543795891 | | |
| 5 | -0.1725443285 | | | |